\documentclass[12pt]{article}
\usepackage{amssymb}

\begin{document}

\bigskip

\bigskip\ 

\begin{center}
\textbf{SEARCHING FOR A CONNECTION BETWEEN}

\textbf{MATROID THEORY AND STRING THEORY}

\bigskip\ 

\smallskip\ 

J. A. Nieto\footnote[1]{%
nieto@uas.uasnet.mx}

\smallskip\ 

\textit{Departamento de Investigaci\'{o}n en F\'{i}sica de la Universidad de
Sonora,}

\textit{83190, Hermosillo Sonora, M\'{e}xico}

\textit{and}

\textit{Facultad de Ciencias F\'{i}sico-Matem\'{a}ticas de la Universidad
Aut\'{o}noma}

\textit{de Sinaloa, 80010, Culiac\'{a}n Sinaloa, M\'{e}xico}

\bigskip\ 

\bigskip\ 

\textbf{Abstract}
\end{center}

We make a number of observations about matter-ghost string phase, which may
eventually lead to a formal connection between matroid theory and string
theory. In particular, in order to take advantage of the already established
connection between matroid theory and Chern-Simons theory, we propose a
generalization of string theory in terms of some kind of Kahler metric. We
show that this generalization is closely related to the Kahler-Chern-Simons
action due to Nair and Schiff. In addition, we discuss matroid/string
connection via matroid bundles and a Schild type action, and we add new
information about the relationship between matroid theory, $D=11$
supergravity and Chern-Simons formalism.

\bigskip\ 

Pacs numbers: 04.60.-m, 04.65.+e, 11.15.-q, 11.30.Ly

May, 2003

\newpage\ 

\noindent \textbf{1. INTRODUCTION}

\smallskip\ 

Although the key principle in M-theory$^{1-3}$ and string theory$^{4}$ is
unknown there is accumulating evidence for the existence of some kind of
duality principle. In fact, duality is the key physical concept that relates
the five known superstring theories in 9+1 dimensions (i.e., nine space and
one time), Type I, Type IIA, Type IIB, Heterotic SO(32) and Heterotic $%
E_{8}\times E_{8}$, which may now be understood as different manifestations
of M-theory. Thus, anticipating the possibility that duality is the basic
principle in M-theory, one may be interested in the mathematical structure
necessary to make sense of such a duality principle. The idea is similar to
the role played by tensor analysis which gives a mathematical sense to the
postulate of relativity ``the laws of physics are the same for every
observer''. In two previous works we proposed the possibility that such a
mathematical structure could be realized through the so called matroid
theory.$^{5}$ Matroid theory, which can be understood as a generalization of
graph theory and matrix theory, has the duality symmetry among its key basic
concepts. In fact, in contrast to graphs in which duality can only be
considered in connection with planar graphs, matroid theory has the
remarkable property that every matroid has a unique dual matroid. As an
example of the importance of the duality property in matroid theory let us
just mention a theorem due to Whitney$^{5}$: if $M_{1},..,M_{p}$ and $%
M_{1}^{^{\prime }},..,M_{p}^{^{\prime }}$ are the components (blocks) of the
matroids $M$ and $M^{\prime }$ respectively, and if $M_{i}^{^{\prime }}$ is
the dual of $M_{i}$ ($i=1,...,p$) then $M^{\prime }$ is dual of $M$ and
conversely, if $M$ and $M^{\prime }$ are dual matroids then $M_{i}^{^{\prime
}}$ is dual of $M_{i}.$ Moreover, in a general context, we have the
remarkable proposition that if a statement $\mu $ in the theory of matroids
has been proved true, then also its dual $\mu ^{\ast }$ is true.

Of course, the question is how to achieve such a relationship between
matroid theory and M-theory. Especially if we do not even know the formal
partition function associated to M-theory. As a first step in this
direction, one may attempt to see if matroid theory is linked somehow to $%
D=11$ supergravity which is one of the manifestations of M-theory. In fact,
it has been shown$^{6}$ that the Fano matroid and its dual are closely
related to Englert's compactification$^{7}$ of $D=11$ supergravity. This
result is physically interesting because it allows a connection between the
fundamental Fano matroid or its dual$^{8}$ and octonions which, at the same
time, are one of the alternative division algebras.$^{9}$ In reference 10,
we made further progress on this program, incorporating matroid theory on
quantum Yang-Mills theory in the context of Chern-Simons action. Our
mechanism was based on a theorem due to Thistlethwaite$^{11}$ which connects
the Jones polynomial for alternating knots with the Tutte polynomial for
graphs. Since Witten$^{12}$ showed that Jones polynomial can be understood
in three dimensional terms through a Chern-Simons formalism, it became
evident that we achieved a bridge between matroid theory and Chern-Simons
formalism.

In this article, we further pursue the idea of relating matroid theory with
M-theory. Since the five fundamental strings are different vacuum limits of
M-theory, it seems natural to try to find first a link between matroid
theory and string theory. In this context there are a number of observations
that indicate that this idea makes sense. First, since Chern-Simons
formalism is closely linked to conformal field theory and Matrix theory,
which in turn are related to string theory, one should expect a connection
of the form: matroid-theory$\rightarrow $Chern-Simons-theory$\rightarrow $%
string-theory. Second, since strings are closely related to knots, which in
turn are related in one to one correspondence to signed graphs, one should
expect a link of the form: matroids$\rightarrow $signed graphs$\rightarrow $%
knots$\rightarrow $strings. Finally, we can in effect combine the two
previous observations in the form: matroids$\rightarrow $signed graphs$%
\rightarrow $knots$\rightarrow $Chern-Simons-formalism$\rightarrow $strings.

In order to achieve our goal, we study the possibility that, in the string
phase of matter-ghost coupling, the world sheet metric and the target
space-time metric become unified in just one metric. We show, in some
detail, that such a unified metric may be a certain kind of Kahler metric.
This observation lead us to consider the Kahler-Chern-Simons action as the
key bridge to connect matroid theory and string theory.

An alternative matroid/string connection can be achieved via the Schild type
action.$^{13}$ In fact, we show that writing the Nambu-Goto action in the
context of Schild formulation, such a connection seems to be a
straightforward extension of the chirotope notion of oriented matroids. We
prove that a local description of the chirotope concept becomes part of the
structure of the Schild action. The relevant structure in this process is
the concept of matroid bundle which has already been developed by the
mathematicians.$^{14-17}$ Finally, we show that in order to complete the
desired connection between matroids and strings it appears necessary to use
the Chern-Simons formulation for strings as proposed by Zaikov$^{18}$.

The plan of this work is as follows. In section 2, we briefly review matroid
theory. In section 3, we closely follow the Ref. 6 adding new information
about the connection between matroid theory and $D=11$ supergravity . In
section 4, we briefly review Ref. 10 and propose a possible extension of the
relation between matroid theory and Witten's partition function for knots.
In section 5, we propose a generalized Polyakov string action with the
property of unifying the world-sheet metric and the target space-time
metric. In section 6, we discuss an alternative matroid/string connection
via the concept of a chirotope of oriented matroids and Schild type action.
Finally, in section 7, we make some final comments.

\bigskip\ \smallskip\ 

\noindent \textbf{2. A BRIEF REVIEW OF MATROID THEORY}

\smallskip\ 

At present matroid theory, also called combinatorial geometry or
pregeometry, can be understood as the combinatorial analogue of K-theory. In
fact, the axioms of K-theory are very similar to the properties achieved
with the Tutte-Gotendiek invariants for matroids. This interpretation
emerged from a great number of contributions from a several mathematicians
since 1935 with the pioneer work of Whitney$^{5}$ on ''Abstract properties
of linear dependence''. In the same year, Birkhoff$^{19}$ established the
connection between simple matroids and geometric lattices. In 1936, MacLane$%
^{20}$ gave an interpretation of matroids in terms of projective geometry.
And an important progress to the subject was given in 1958 by Tutte$^{8}$
who introduced the concept of homotopy for matroids. The fascination of this
subject among the combinatorial mathematicians can be appreciated from the
large body of information about matroid theory. In fact, there is a large
number of books about matroid theory. For background information on this
subject the reader should consult Oxley$^{21}$ and Welsh.$^{22}$ We also
recommend the books of Wilson,$^{23}$ Kung$^{24}$ and Ribnikov.$^{25}$

It is known that in graph theory only planar graphs have an associated dual
graph. For instance, the Kuratowski theorem assures that the complete graph $%
K_{5}$ and the bipartita graph $K_{3,3}$ which are not planar do not have an
associated dual graph. In a sense, matroid theory arose as an attempt to
solve this lack of duality symmetry. The attractive feature is that in
matroid theory every matroid has an associated unique dual matroid. In
particular the matroid associated to $K_{5}$, let us say $M(K_{5}),$ has a
dual $M^{\ast }(K_{5})$. The important aspect is that $M^{\ast }(K_{5})$ is
not graphic, that is, it can not be represented by a graph. This is, of
course, an indication that matroid theory is a generalization of graph
theory. Therefore, the great advantage of matroid theory is that it provides
us with a mathematical structure in which the concepts of duality of planar
graphs is extended to graphs that are not planar.

Another interesting aspect that motivates the subject is that linear
dependence in algebra can be understood as a particular case of matroid
theory. In fact, matroid theory leads to matroids that are not even
representable by a finite set of vectors in a vector space or by matrices,
extending the concept of orthogonality in vector spaces. Summarizing, we can
say that by extending the concept of duality in vector spaces and planar
graphs, matroid theory accomplishes a generalization of both graph theory
and matrix theory.

Mathematically, a matroid is defined as follows: a matroid $M$ is a pair $%
(E,I)$, where $E,$ called the ground set, is a non-empty finite set, and $I$
is a non-empty collection of subsets of $E$ satisfying the following two
properties:

$(I$ $\mathit{i)}$\textit{\ }any subset of an independent set is independent.

$(I\mathit{\ ii)}$\textit{\ }if $K$ and $J$ are independent sets with $%
K\subseteq $ $J$, then there is an element $e$ contained in $J$ but not in $%
K $, such that $K\cup \{e\}$ is independent.

Members of $I$ are called independent sets of $M$; other sets are called
dependent. Therefore, the definition itself of a matroid divides all
possible subsets of $E$ in two types: independent and dependent subsets.
Thus, we see that, even from the beginning, matroids have the dual structure
independent-dependent. From this point of view, it is not a surprise to find
eventually that every matroid has an associated dual matroid.

A base is defined to be any maximal independent set. Similarly, the minimal
dependent set is called a circuit. By repeatedly using the property $(I%
\mathit{ii)}$\textit{\ }it is straightforward to show that any two bases
have the same number of elements.

An alternative definition of a matroid in terms of bases is as follows:

A matroid $M$ is a pair $(E,\mathcal{B})$, where $E$ is a non-empty finite
set and $\mathcal{B}$ is a non-empty collection of subsets of $E$ (called
bases) satisfying the following properties:

$(\mathcal{B}$ $\mathit{i)}$\textit{\ }no base properly contains another
base;

$(\mathcal{B}$ $\mathit{ii)}$ if $B_{1}$ and $B_{2}$ are bases and if $b$ is
any element of $B_{1},$ then there is an element $g$ of $B_{2}$ with the
property that $(B_{1}-\{b\})\cup \{g\}$ is also a base.

A matroid can also be defined in terms of circuits:

A matroid $M$ is a pair $(E$, $C)$, where $E$ is a non-empty finite set, and 
$C$ is a non-empty collection of subsets of $E$ (called circuits) satisfying
the following properties.

$(C$ $\mathit{i})$ no circuit properly contains another circuit;

$(C$ $\mathit{ii)}$\textit{\ }if $C_{1}$ and $C_{2}$ are two distinct
circuits each containing an element $c$, then there exists a circuit in $%
C_{1}\cup C_{2}$ which does not contain $c$.

If we start with any of the three definitions then one finds that the other
two follow as theorems. For example, it is possible to prove that $(I\mathit{%
)}$ implies $(\mathcal{B})$ and $(C)$. In other words, these three
definitions are equivalent. There are other definitions also equivalent to
these three, but for the purpose of this work it is not necessary to
consider all of them.

As we noticed previously, even from the initial structure of a matroid
theory we find relations such as independent-dependent structure which
suggests duality. The dual of $M$, denoted by $M^{\ast },$ is defined as a
pair $(E,\mathcal{B}^{\ast }),$ where $\mathcal{B}^{\ast }$ is a non-empty
collection of subsets of $E$ formed with the complements of the bases of $M$%
. An immediate consequence of this definition is that every matroid has a
dual and this dual is unique. It also follows that the double-dual $M^{\ast
\ast }$ is equal to $M$. Moreover, if $S$ is a subset of $E$, then the size
of the largest independent set contained in $S$ is called the rank of $S$
and is denoted by $\rho (S)$. If $M=M_{1}+M_{2}$ and $\rho (M)=\rho
(M_{1})+\rho (M_{2})$ we shall say that $M$ is separable. Any maximal
non-separable part of $M$ is a block of $M$. An important theorem due to
Whitney$^{5}$ is that if $M_{1},..,M_{p}$ and $M_{1}^{^{\prime
}},..,M_{p}^{^{\prime }}$ are the blocks of the matroids $M$ and $M^{\prime
} $ respectively, and if $M_{i}^{^{\prime }}$ is the dual of $M_{i}$ $%
(i=1,...,p)$. Then $M^{\prime }$ is dual of $M$. Conversely, let $M$ and $%
M^{\prime }$ be dual matroids, and let $M_{1},..,M_{p}$ be blocks of $M$.
Let $M_{1}^{^{\prime }},..,M_{p}^{^{\prime }}$ be the corresponding
submatroids of $M^{\prime }$. Then $M_{1}^{^{\prime }},..,M_{p}^{^{\prime }}$
are the blocks of $M^{\prime }$, and $M_{i}^{^{\prime }}$ is dual of $M_{i}.$

Over the last years matroid theory has been growing very rapidly. There are
already well established formalisms for oriented matroids$^{26}$ and bias
matroids.$^{27}$ The former can be understood as a generalization of
oriented graphs and the latter as an extension of signed graphs. In each one
of these branches of matroid theory there are very interesting theorems and
results, some of which we shall mention in the next sections.

\bigskip\ \smallskip\ 

\noindent \textbf{3. MATROID THEORY AND SUPERGRAVITY}

\smallskip\ 

Here, we briefly review the main results of Ref. 6 and add some new
observations. In Ref. 6 we showed that the Fano matroid $F_{7}$may be
connected with octonions which, in turn, are related to the Englert's
compactification of $D=11$ supergravity.

A Fano matroid $F_{7}$ is the matroid defined on the ground set

\begin{equation}
E=\{1,2,3,4,5,6,7\}  \label{1}
\end{equation}
whose bases are all those subsets of $E$ with three elements except $%
f_{1}=\{1,2,3\},$ $f_{2}=\{5,1,6\},$ $f_{3}=\{6,4,2\},$ $f_{4}=\{4,3,5\},$ $%
f_{5}=\{4,7,1\},$ $f_{6}=\{6,7,3\}$ and $f_{7}=\{5,7,2\}$. The circuits of
the Fano matroid are precisely these subsets and its complements. It follows
that these circuits define the dual $F_{7}^{\ast }$ of the Fano matroid.

Let us write the set $E$ in the form\textit{\ }$E$=$%
\{e_{1},e_{2},e_{3,}e_{4,}e_{5},e_{6},e_{7}\}$. Thus, the subsets used to
define the Fano matroid now become $f_{1}=\{e_{1},e_{2},e_{3}\}$, $%
f_{2}=\{e_{5},e_{1},e_{6}\}$, $f_{3}=\{e_{6},e_{4},e_{2}\}$, $%
f_{4}=\{e_{4},e_{3},e_{5}\}$, $f_{5}=\{e_{4},e_{7},e_{1}\}$, $%
f_{6}=\{e_{6},e_{7},e_{3}\}$ and $f_{7}=\{e_{5},e_{7},e_{2}\}$. The key idea
in Ref. 6 was to identify the quantities $e_{i},$ where $i=1,...$,$7,$ with
the octonionic imaginary units. Specifically, we write an octonion $q$ in
the form 
\[
q=q_{0}e_{0}+q_{1}e_{1}+q_{2}e_{2}+q_{3}e_{3}+q_{4}e_{4}+q_{5}e_{5}+q_{6}e_{6}
+q_{7}e_{7}, 
\]
where $q_{0}$ and $q_{i}$ are real numbers. Here, $e_{0}$ denotes the
identity. The product of two octonions can be obtained from the formula:

\begin{equation}
e_{i}e_{j}=-\delta _{ij}+\psi _{ij}^{k}e_{k},  \label{2}
\end{equation}
where $\delta _{ij}$ is the Kronecker delta and $\psi _{ijk}=$ $\psi
_{ij}^{l}\delta _{lk}$ is the fully antisymmetric structure constants, with $%
i,j,k=1,...,7$. By taking the $\psi _{ijk}$ equals $1$ or $-1$ for each one
of the seven combinations $f_{i}$ we may derive all the values of $\psi
_{ijk}$.

The octonion (Cayley) algebra is not associative, but alternative. This
means that the basic associator of any three imaginary units is

\begin{equation}
\langle e_{i},e_{j},e_{k}\rangle
=(e_{i}e_{j})e_{k}-e_{i}(e_{j}e_{k})=\varphi _{ijkm}e_{m},  \label{3}
\end{equation}
where $\varphi _{ijkl}$ is a fully antisymmetric four index tensor. It turns
out that $\varphi _{ijkl}$ and $\psi _{ijk}$ are related by the expression

\begin{equation}
\varphi _{ijkl}=(1/3!)\epsilon _{ijklmnr}\psi _{mnr},  \label{4}
\end{equation}
where $\epsilon _{ijklmnr}$ is the completely antisymmetric Levi-Civita
tensor, with $\epsilon _{12...7}=1$. It is interesting to observe that given
the numerical values $f_{i}$ for the indices of $\psi _{mnr}$ and using (4)
we get the other seven subsets of $E$ with four elements of the dual Fano
matroid $F_{7}^{\ast }.$ For instance, if we take $f_{1}$ then we have $\psi
_{123}$ and (4) gives $\varphi _{4567}$ which leads to the circuit subset $%
\{4,5,6,7\}$ of $F_{7}^{\ast }.$

Therefore, this shows that the Fano matroid and its dual are closely related
to octonions which at the same time are an essential part of the Englert's
solution of absolute parallelism on $S^{7}$ of $D=11$ supergravity. It is
important to mention that the Fano matroid is the only minimal binary
irregular matroid. Just as octonions are central mathematical objects in
division algebras, this property makes the Fano matroid a central
mathematical object in matroid theory. $D=11$ supergravity is on the other
hand an important physical structure in M-theory. Therefore we have here a
link between three apparently unrelated important objects in its own field:
Fano matroid (matroid theory) $\leftrightarrow $octonions (algebra) $%
\leftrightarrow D=11$ supergravity (unify fundamental physics).

We would like to make some further observations about the link between the
Fano matroid and octonions. Consider the subsets $h_{1}=\{v_{1},v_{2},v_{3}%
\} $, $h_{2}=\{v_{5},v_{1},v_{6}\}$, $h_{3}=\{v_{6},v_{2},v_{4}\}$, $%
h_{4}=\{v_{4},v_{3},v_{5}\}$, $h_{5}=\{v_{4},v_{7},v_{1}\}$, $%
h_{6}=\{v_{6},v_{7},v_{3}\}$ and $h_{7}=\{v_{5},v_{7},v_{2}\}.$ If we
identify $v_{i},$ where $i=1,...$,$7,$ of these subsets with the columns of
the matrix

\begin{equation}
A=\left( 
\begin{array}{ccccccc}
1 & 0 & 1 & 0 & 1 & 0 & 1 \\ 
1 & 1 & 0 & 0 & 0 & 1 & 1 \\ 
0 & 1 & 1 & 1 & 0 & 0 & 1
\end{array}
\right) ,  \label{5}
\end{equation}
we notice that the matrix $A$ provides a representation (or realization) of
the Fano matroid $F_{7}$. Now, suppose that the Fano matroid is extended to
a structure in which the sets $\{v_{i},v_{j},v_{k}\}$ corresponding to the $%
h_{i}$ are replaced by the completely antisymmetric object

\begin{equation}
(v_{i},v_{j},v_{k})=-(v_{i},v_{k},v_{j})=-(v_{j},v_{i},v_{k}).  \label{6}
\end{equation}
For instance, we may replace $h_{1}=\{v_{1},v_{2},v_{3}\}$ by $\hat{h}%
_{1}=(v_{1},v_{2},v_{3})$. Specifically, we define the extended Fano matroid 
$\hat{F}_{7}$ as the pair $(E,\mathcal{\hat{B}})$ in which $\mathcal{\hat{B}}
$ is the set of three elements $(v_{i},v_{j},v_{k})$ except the completely
antisymmetric quantities $\hat{h}_{1}=(v_{1},v_{2},v_{3})$, $\hat{h}%
_{2}=(v_{5},v_{1},v_{6})$, $\hat{h}_{3}=(v_{6},v_{2},v_{4})$, $\hat{h}%
_{4}=(v_{4},v_{3},v_{5})$, $\hat{h}_{5}=(v_{4},v_{7},v_{1}),$ $\hat{h}%
_{6}=(v_{6},v_{7},v_{3})$ and $\hat{h}_{7}=(v_{5},v_{7},v_{2})$. The
generalization from $F_{7}$ to $\hat{F}_{7}$ is very similar to the
transition from graphs to digraphs (or oriented graphs)\ in which the edges
of the original graph, let us say $\{a,b\},$ are changed to an ordering set, 
$(a,b)=-(b,a)$. The important point is that if there exists such a
transition between $F_{7}$ and $\hat{F}_{7}$ then we discover that $\hat{F}%
_{7}$ almost determine completely the octonion algebra, essentially because $%
(v_{i},v_{j},v_{k})$ for the different $h_{i}$ become closely related to the
structure constants $\psi _{mnr}$ associated to octonions. In fact, there is
an extension of matroid theory which seems to be what these observations
suggest for the Fano matroid, namely oriented matroids.$^{26}$

In order to define oriented matroids it is necessary to define first what
signed circuits are. A signed circuit $X$ is a circuit with the partition $%
(X^{+},X^{-})$ into two sets: $X^{+}$ the set of positive elements of $X$,
and $X^{-}$ its set of negative elements.

An oriented matroid $\mathcal{M}$ is a pair $(E,\mathcal{C})$, where $E$ is
a non-empty finite set, and $\mathcal{C}$ is a non-empty collection of
subsets of $E$ (called signed circuits) satisfying the following properties.

($\mathcal{C}$ \textit{i}) no circuit properly contains another circuit.

($\mathcal{C}$ \textit{ii) }if $\mathcal{C}_{1}$ and $\mathcal{C}_{2}$ are
two distinct signed circuits, $\mathcal{C}_{1}\neq $-$\mathcal{C}_{2},$ and $%
c\in \mathcal{C}_{1}^{+}\cap \mathcal{C}_{2}^{-}$ then there exists a third
circuit $\mathcal{C}_{3}\in \mathcal{C}$ with $\mathcal{C}_{3}^{+}\subseteq (%
\mathcal{C}_{1}^{+}\cap \mathcal{C}_{2}^{+})\setminus \{c\}$ and $\mathcal{C}%
_{3}^{-}\subseteq (\mathcal{C}_{1}^{-}\cap \mathcal{C}_{2}^{-})\setminus
\{c\}.$

It is not difficult to see that by forgetting signs, this definition of
oriented matroids reduces to the definition of ordinary (non-oriented)
matroids.

An alternative but equivalent way to define an oriented matroid is as
follows: An oriented matroid $\mathcal{M}$ is a pair $(E,\chi ),$ where $E$
is a non-empty finite set and $\chi $ (called chirotope) is a mapping $%
E^{r}\rightarrow \{-1,0,1\},$ with $r$ the rank on $E,$ satisfying the
following properties.

$(\chi i)$ $\chi $ is not identically zero

$(\chi ii)\chi $ is alternating

$(\chi iii)$ for all $x_{1},x_{2},...,x_{r}$ and $y_{1},y_{2},...,y_{r}$
such that

\begin{equation}
\chi (x_{1},x_{2},...,x_{r})\chi (y_{1},y_{2},...,y_{r})\neq 0  \label{7}
\end{equation}
there exists an $i\in \{1,2,3,4,5,6,7\}$ such that

\begin{equation}
\chi (y_{i},x_{2},...,x_{r})\chi
(y_{1},y_{2},...,y_{i-1},x_{1},y_{i+1,}...,y_{r})=\chi
(x_{1},x_{2},...,x_{r})\chi (y_{1},y_{2},...,y_{r}).  \label{8}
\end{equation}
For a vector configuration $\chi $ can be identified as

\begin{equation}
\chi (i_{1},...,i_{r})\equiv sign\det (v_{i_{1}},...,v_{i_{r}})\in \{-1,0,1\}
\label{9}
\end{equation}
and $(\chi iii)$ turns out to be related to the Grassmann-Plucker relation.

Returning to the case of the Fano matroid it is tempting to identify $h_{i}$
with the chirotope 
\begin{equation}
\chi (i_{1},i_{2},i_{3})=sign\det (v_{i_{1}},v_{i_{2}},v_{i_{3}}).
\label{10}
\end{equation}
But, in Ref. 28 it is noted that the Fano matroid is not orientable.
Specifically, one can verify that the Fano matroid does not satisfy the
property $(\chi iii)$. Nevertheless, it is interesting to observe that one
may write the formula$^{29}$

\begin{equation}
\psi _{i_{1}i_{2}i_{3}}+\chi (i_{1},i_{2},i_{3})=C_{i_{1}i_{2}i_{3}},
\label{11}
\end{equation}
where $C_{i_{1}i_{2}i_{3}}\in \{-1,1\}$ may be identified with the uniform
matroid $U_{3,7}$ which is an excluded minor for $GF(5)$-representability,
where $GF(q)$ denotes a finite field of order $q.$ In this sense the Fano
matroid and the octonions look as complementary concepts of the oriented
uniform matroid $M(U_{3,7})$ structure.

It is worth remarking that the structure of $\hat{F}_{7}$ not necessarily
corresponds, in a straightforward way, to oriented matroids for the
following observations. An important problem in matroid theory is to see
which matroids can be mapped into a set of vectors in a vector space over a
given field. When such a map exists we are speaking about a coordinatization
(or representation) of the matroid over the field. A matroid which has a
coordinatization over $GF(2)$ is called binary. Furthermore, a matroid which
has a coordinatization over every field is called regular. It turns out that
regular matroids are of fundamental importance in matroid theory, among
other things, because they play a similar role as planar graphs in graph
theory.$^{23}$ It is known that a graph is planar if and only if it contains
no subgraph homeomorphic to $K_{5}$ or $K_{3,3}$. The analogue of this
theorem for matroids was proved by Tutte.$^{8}$ In fact, Tutte proved that a
matroid is regular if and only if it is binary and has no minor isomorphic
to the Fano matroid or the dual of this.

The important point is that an algebra, like the octonion algebra, is a
vector space with an additional multiplicative operation. If it could be
possible to identify this additional product with a kind of rule for the
bases of $\mathcal{B}$ in a given matroid then we could speak about a
representation of a matroid (with this additional product) in terms of an
algebra instead of just the corresponding vector space. At present, we have
not been able to find in the literature this kind of structure for matroids.
But it seems to us that our identification of $\hat{F}_{7}$ with octonions
may provide an example of matroids associated with an algebra rather than
with just the corresponding vector space.

\bigskip\ \smallskip\ 

\noindent \textbf{4. MATROID THEORY AND CHERN-SIMONS THEORY}

\smallskip\ 

Here, we shall briefly review the main results of Ref. 10 about the
connection between matroid theory and Chern-Simons theory and make
additional comments. For this purpose let us introduce the Witten's
partition function

\begin{equation}
Z(L,k)=\int DA\exp (S_{cs})\prod\limits_{r=1}^{n}W(L_{r},\rho _{r}),
\label{12}
\end{equation}
where $S_{CS}$ is the Chern-Simons action

\begin{equation}
S_{CS}=\frac{k}{2\pi }\int_{M^{3}}Tr(A\wedge dA+\frac{2}{3}A\wedge A\wedge
A),  \label{13}
\end{equation}
and $W(C_{i},\rho _{i})$ is the Wilson line

\begin{equation}
W(L_{r},\rho _{r})=Tr_{\rho _{r}}P\exp (\smallint
_{L_{r}}A_{i}^{a}T_{a}dx^{i}).  \label{14}
\end{equation}
Here, $A=A_{i}^{a}T_{a}dx^{i}$, with $T_{a}$ the generators of the Lie
algebra of $G$ and the symbol $P$ means the path-ordering along the knots $%
L_{r}.$ If we choose $M^{3}=S^{3},$ $G=SU(2)$ and $\rho _{r}=C^{2}$ for all
the link components then the Witten's partition function (12) reproduces the
Jones polynomial

\begin{equation}
Z(L,k)=V_{L}(t),  \label{15}
\end{equation}
where

\begin{equation}
t=e^{\frac{2\pi i}{k}}  \label{16}
\end{equation}
and $V_{L}(t)$ denotes the Jones polynomial satisfying the skein relation;

\begin{equation}
t^{-1}V_{L_{+}}-tV_{L_{-}}=(\sqrt{t}-\frac{1}{\sqrt{t}})V_{L_{0}},
\label{17}
\end{equation}
where $L_{+},L_{-}$ and $L_{0}$ are the standard notation for overcrossing,
undercrossing and zero crossing.

On the other hand, Thistlethwaite$^{11}$ showed that if $L$ is an
alternating link and $\mathit{G}(L)$ the corresponding planar graph, then
the Jones polynomial $V_{L}(t)$ is equal to the Tutte polynomial $T_{\mathit{%
G}}(-t,-t^{-1})$ up to a sign and a factor power of $t.$ Specifically, we
have

\begin{equation}
V_{L}(t)=(-t^{\frac{3}{4}})^{w(L)}t^{\frac{-(\rho -n)}{4}}T_{\mathit{G}%
}(-t,-t^{-1}),  \label{18}
\end{equation}
where $w(L)$ is the writhe and $\rho $ and $n$ are the rank and the nullity
of $\mathit{G,}$ respectively$.$ Here, $V_{L}(t)$ is the Jones polynomial of
alternating link $L.$ The Tutte polynomial associated to each graph $\mathit{%
G}$ is a polynomial $T_{\mathit{G}}(x,x^{-1})$ with the property that if $%
\mathit{G}$ is composed solely of isthmus and loops then $T_{\mathit{G}%
}(x,x^{-1})=x^{I}x^{-l},$ where $I$ is the number of isthmuses and $l$ is
the number of loops. The polynomial $T_{\mathit{G}}$ satisfies the skein
relation

\begin{equation}
T_{\mathit{G}}=T_{\mathit{G}^{^{\prime }}}+T_{\mathit{G}^{^{\prime \prime
}}},  \label{19}
\end{equation}
where $\mathit{G}^{\prime }$ and $\mathit{G}^{\prime \prime }$ are obtained
by delating and contracting respectively an edge that is neither a loop nor
an isthmus of $\mathit{G}$.

On the other hand, a theorem due to Tutte allows to compute $T_{\mathit{G}%
}(-t,-t^{-1})$ from the maximal trees of $\mathit{G}$. In fact, Tutte proved
that if $\mathcal{B}$ denotes the set of maximal trees in a graph $\mathit{G}
$, $i(\mathcal{B})$ denotes the number of internally active edges in $%
\mathit{G,}$ and $e(B)$ refers to the number of the externally active edges
in $\mathit{G}$ (with respect to a given maximal tree $B\in \mathcal{B}$)
then the Tutte polynomial is given by the formula

\begin{equation}
T_{\mathit{G}}(-t,-t^{-1})=\sum_{B\subseteq \mathcal{B}}x^{i(B)}x^{-e(B)},
\label{20}
\end{equation}
where the sum is over all elements of $\mathcal{B}$.

The important point is that the Tutte polynomial $T_{\mathit{G}}(-t,-t^{-1})$
computed according to (20) uses the concept of a graphic matroid $M(\mathit{%
G)}$ defined as the pair $(E,\mathcal{B})$, where $E$ is the set of edges of 
$\mathit{G}$. This remarkable connection between the Tutte polynomial and a
matroid allows in fact a relation between the partition function $Z(L,k)$
given in (12) and matroid theory. This is because according to (18) the
Tutte polynomial $T_{\mathit{G}}(-t,-t^{-1})$ is related to the Jones
polynomial $V_{L}(t)$ which in turn according to (15) is connected to the
partition function $Z(L,k)$. Specifically, for $M^{3}=S^{3},$ $G=SU(2)$, $%
\rho _{r}=C^{2}$ for all alternating link components of $L$, we have the
relation

\begin{equation}
Z(L,k)=V_{L}(t)=(-t^{\frac{3}{4}})^{w(L)}t^{\frac{-(\rho -n)}{4}}T_{\mathit{G%
}}(-t,-t^{-1}).  \label{21}
\end{equation}
Thus, the matroid $(E,\mathcal{B})$ used to compute $T_{\mathit{G}%
}(-t,-t^{-1})$ can be associated not only to $V_{L}(t),$ but also to $%
Z(L,k). $ Therefore, we have found a bridge which links the matroid
formalism $(E,\mathcal{B})$ and the partition function $Z(L,k).$ This may
allow to bring many concepts of matroid theory to fundamental physics and
conversely, different results in fundamental physics may be used as an
inspiration to further develop matroid theory. As an example of the former
remark let us just mention how the duality concept in matroid theory can be
used as a symmetry of $Z(L,k).$

First of all, it is known that in matroid theory the concept of duality is
of fundamental importance. For example, there is a remarkable theorem that
assures that every matroid has a dual. So, the question arises about what
are the implications of this theorem in Chern-Simons formalism. In order to
address this question let us first make a change of notation $T_{\mathit{G}%
}(-t,-t^{-1})\rightarrow T_{M(\mathit{G)}}(t)$ and $Z(L,k)\rightarrow Z_{M(%
\mathit{G)}}(k).$ The idea of this notation is to emphasize the connection
between matroid theory, Tutte polynomial and Chern- Simons partition
function. Consider the planar dual graph $\mathit{G}^{\ast }$ of $\mathit{G}$%
. In matroid theory we have $M(\mathit{G}^{\ast })$ =$M^{\ast }(\mathit{G)}$%
. Therefore, the duality property of the Tutte polynomial

\begin{equation}
T_{\mathit{G}}(-t,-t^{-1})=T_{\mathit{G}^{\ast }}(-t^{-1},-t)  \label{22}
\end{equation}
can be expressed as

\begin{equation}
T_{M(\mathit{G)}}(t)=T_{M^{\ast }(\mathit{G)}}(t^{-1})  \label{23}
\end{equation}
and consequently from (15) and (18) we discover that for the partition
function $Z_{M(\mathit{G)}}(k)$ we should have the dual property

\begin{equation}
Z_{M(\mathit{G)}}(k)=Z_{M^{\ast }(\mathit{G)}}(-k).  \label{24}
\end{equation}

As a second example let us first mention another theorem due to Withney:$%
^{5} $ If $M_{1},..,M_{p}$ and $M_{1}^{^{\prime }},..,M_{p}^{^{\prime }}$
are the components (or blocks) of the matroids $M$ and $M^{\prime }$
respectively, and if $M_{i}^{^{\prime }}$ is the dual of $M_{i}$ $%
(i=1,...,p) $. Then $M^{\prime }$ is dual of $M$. Conversely, let $M$ and $%
M^{\prime }$ be dual matroids, and let $M_{1},..,M_{p}$ be components of $M$%
. Let $M_{1}^{^{\prime }},..,M_{p}^{^{\prime }}$ be the corresponding
submatroids of $M^{\prime }$. Then $M_{1}^{^{\prime }},..,M_{p}^{^{\prime }}$
are the components of $M^{\prime }$, and $M_{i}^{^{\prime }}$ is dual of $%
M_{i}.$ Thus, according to (24) we find that

\begin{equation}
Z_{M_{i}(\mathit{G}_{i}\mathit{)}}(k)=Z_{M_{i}^{\prime }(\mathit{G}_{i}%
\mathit{)}}(-k)  \label{25}
\end{equation}
if and only if

\begin{equation}
Z_{M(\mathit{G)}}(k)=Z_{M^{\prime }(\mathit{G)}}(-k),  \label{26}
\end{equation}
where $\mathit{G}_{i}$ are the components or blocks of $\mathit{G.}$

Our discussion has been, so far, based on alternating links $L.$ This kind
of links is an important, but relatively small subclass of links. In fact,
there is a one to one correspondence between links and signed graphs and a
link is alternating if the signed graph representation has all edges with
the same sign. Therefore, in order to generalize the procedure it turns
necessary to have a generalized Thustlethwaite's $^{11}$ theorem for any
signed graph not just for those of the same sign. Fortunately,
Thustlethwaite himself,$^{11}$ and later Kauffman,$^{30}$ precisely
generalized the original Thustlethwaite's theorem for planar unsigned graphs.

Theorem: Let $G$ be a planar signed graph. Let $K(G)$ be the knot/link
diagram corresponding to $G$. Then $\langle K(G)\rangle =T_{G}(A,B,x,y)$.
The bracket polynomial for knots and links is a specialization of the
generalized Tutte polynomial for signed graphs.

Here, $A,B,$and $d$ are commuting variables associated to the link. $A$ and $%
B$ correspond to $A$-channel, $B$-channel respectively, while the parameter $%
d$ is used as a factor of normalization in order to make $T_{G}(A,B,d)$
invariant under the Reidemeister moves II and III.

Furthermore, Kauffman showed that $T_{G}=T_{G}(A,B,x,y)$ has a spanning tree
expansion of the form

\begin{equation}
T_{G}=\sum_{H\subseteq \mathcal{B}}\Lambda (H),  \label{27}
\end{equation}
where $\Lambda (H)$ denotes the product of the contribution of the edges of $%
G$ relative activities of the maximal trees $H$ in $G$.

In principle, since up to a normalized factor, measuring the orientability
of the link, $\langle K(G)\rangle \leftrightarrow CS,$ in order to find a
generalization of our procedure we need to relate $T_{G}$ with matroid
theory. A $T_{G}\longleftrightarrow matroids$ connection is given by (27) in
the sense that the sum is over all maximal trees $H$ in $G$. Notice,
however, that the maximal trees $H$ are associated to the underlying graph
(without signs) of the signed graph and not to the signed graph itself.

It is known that matroids associated to signed graphs are called bias
matroids.$^{27}$ It turns out that bias matroids are interesting by
themselves, but unfortunately the subject about this kind of matroids has
not been developed for our purpose and it appears that many of the
interesting properties of ordinary matroids are lost. Nevertheless, the idea
of writing $T_{G}$ as a sum over bias matroids seems interesting and
deserves further study.

It may help to mention in this direction that Crapo$^{31}$ proposed an
alternative possibility to write $T_{G}$ as a sum over all spanning subsets
of $E,$ rather than over maximal trees. This idea is motivated from the
observation that in this case the rank function $\rho $ becomes an important
concept and can be used to generalize $T_{G}$ to matroid theory. A
generalization for signed graphs of the Crapo's polynomial has been proposed
by Murasugi$^{32}$ and by Shwarzler and Welsh.$^{33}$ Let us briefly mention
these two polynomials.

Murasugi introduced the following polynomial. Let $\Gamma (r,s)$ denote the
set of all spanning subgraph $S$ of $G.$ Then $T_{G}(x,y,z)$ is defined by

\begin{equation}
T_{G}(x,y,z)=\sum_{k,\rho }\left\{ \sum_{S\in \Gamma
(r,s)}x^{P(S)-N(S)}\right\} y^{k(S)-1}z^{\mid S\mid -\rho (S)},  \label{28}
\end{equation}
where $P(S)$ and $N(S)$ denote the number of positive and negative edges in $%
S$ respectively. It is interesting to note that $\beta _{0}=k=r+1$ and $%
\beta _{1}=n=\mid S\mid -\rho (S)$, where $n$ is the nullity and $\beta _{i}$
denotes the $i$ Betti number of $S$ as a $1-$complex$.$ Although this
polynomial uses the rank and the nullity concepts, the fact that the sum is
over all spanning subgraphs $\Gamma $ means that $T_{G}(x,y,z)$ is also
applied only to the underlying unsigned graph $G$ associated to the signed
graph. Furthermore, the Murasugi polynomial does not have a direct relation
with the Kauffmann polynomial.

On the other hand, Shwarzler and Welsh$^{33}$ proved that the Kauffmann
polynomial associated to a link $L$ can be expressed in terms of the
associated signed graph $G(L)$ as follows

\begin{equation}
T_{G}(A,B)=A^{\mid E^{-}\mid -\mid E^{+}\mid }(-A^{2}-A^{-2})^{\rho
(G)}\sum_{S\subseteq E}A^{4(\rho (S)-\mid S^{-}\mid )}B^{\rho (G)+\mid S\mid
-2\rho (S)},  \label{29}
\end{equation}
where $B=-A^{4}-1$ and for any subset $S\subseteq E(G)$, $S^{+}$ and $S^{-}$
denote the positive and negative signed part respectively. It is important
to remark that Shwarzler and Welsh showed that (29) is a specialization of a
more general polynomial for signed matroids. In fact, Shwarzler and Welsh
proposed an eight variables polynomial which contains as specialization not
only the Kauffman bracket polynomial but also the Tutte polynomial of a
matroid, the partition function of the anisotropic Ising model and the
Kauffman-Murasugi polynomial of signed graphs (for further details see Ref.
33).

\bigskip\ \smallskip\ 

\noindent \textbf{5. MATROID THEORY AND STRING THEORY}

\smallskip\ 

In the literature,$^{34-36}$ several attempts have been done to connect
Chern-Simons formalism with string theory. One of the most interesting$^{37}$
comes from the idea that at some level the decoupling of ghost and matter
does not hold. In this case, matter fields and ghosts become mixed and the
standard string theory should be replaced by some kind of topological string
theory.$^{38}$ It has been shown$^{35}$ that some topological string
theories perturbatively coincide with Chern-Simons theory. So, in this sense
Chern-Simons theory is equivalent to topological string theory. However, the
problem arises when it is attempted to relate Chern-Simons theory with
fundamental strings. In fact it has been shown$^{38}$ that in the pure
Chern-Simons formalism there are not enough degrees of freedom to reproduce
not only the induced gravity but the toroidal compactification of heterotic
string.

These observations are, of course, important in order to find a matroid
theory and string theory connection and eventually M-theory connection. In
the previous section we explained a matroid theory-Chern-Simons theory
relation via Tutte and Jones polynomials. It is clear then that what we
should look for is some kind of generalization of fundamental strings which
may provide the bridge between fundamental strings and topological strings.

The generalized fundamental strings could be the topological membrane$^{39}$
itself , but this is likely to be reduced to the topological strings rather
than to fundamental strings. Another possibility is the membrane theory or
any other p-brane$^{40}$, but it has been shown$^{41}$ that through double
dimensional reduction these are reduced to fundamental strings rather than
to topological strings. So, although there is the hope that at some level 3D
topological field theory may lead to fundamental strings, the correct
formulation of such a theory is at present unknown.

In this section, we propose an alternative generalization of fundamental
strings which seems closer to our purpose than the already known alternative
of topological membranes or p-branes.

The idea comes from the observation that in the Polyakov type action the
world sheet metric and the target space-time metrics are decoupled. But it
seems natural to think that at a more fundamental level when ghost and
matter fields are mixed the decoupling between such two metrics is no longer
true. Therefore the desired generalization of string theory must be based on
a unified metric of the world sheet and target space-time metrics.

Let us clarify these observations. For this purpose, let us first consider
the Polyakov action

\begin{equation}
S=\frac{1}{2}\int d^{2}\xi \sqrt{-g}g^{ab}(\xi )\partial _{a}x^{\mu
}\partial _{b}x^{\nu }G_{\mu \nu }(x),  \label{30}
\end{equation}
where $g_{ab}(\xi )$ and $G_{\mu \nu }(x),$ with $\mu ,\nu =1,...,D,$ are
the world sheet metric and the target space-time metric, respectively. We
observe from (30) that the two metrics $g_{ab}(\xi )$ and $G_{\mu \nu }(x)$
play very different roles; $g_{ab}(\xi )$ determines the world sheet metric
swept out by the string in its dynamical evolution, while $G_{\mu \nu }(x)$
determines the background metric where the string is moving. Therefore,
classically $g_{ab}(\xi )$ and $G_{\mu \nu }(x)$ are unrelated. However,
this is no longer true at the quantum level. For instance, it is well known
that in a consistent quantum string theory $g_{ab}(\xi )$ plays an essential
role to fix the size of the matrix $G_{\mu \nu }(x)$: $D=26$ in the bosonic
case. This kind of relation between $g_{ab}(\xi )$ and $G_{\mu \nu }(x)$ is,
however, in a certain sense superficial because in the critical dimension 26
matter fields decouple from the corresponding ghost with associated central
charge $c=-26.$ The important observation is that, as it was mentioned in
the introduction, at a deeper level the decoupling between matter fields and
ghost must be no longer true and therefore one should expect that in such a
case there must be a unified framework for the two metrics $g_{ab}(\xi )$
and $G_{\mu \nu }(x).$

Consider the line element

\begin{equation}
ds^{2}=G_{(\hat{\mu}\hat{\nu})}(x^{\hat{\alpha}})dx^{\hat{\mu}}\otimes dx^{%
\hat{\nu}},  \label{31}
\end{equation}
where the indices $\hat{\mu},$ $\hat{\nu}$ run from $1$ to $2D$, the symbol $%
\otimes $ means tensor product and $G_{(\hat{\mu}\hat{\nu})}=G_{(\hat{\nu}%
\hat{\mu})}$. Suppose that (31) can be written as

\begin{equation}
ds^{2}=G_{(\mu \nu )}(x^{\hat{\alpha}})dx^{\mu }\otimes dx^{\nu }+G_{(\mu
\nu )}(x^{\hat{\alpha}})dy^{\mu }\otimes dy^{\nu }.  \label{32}
\end{equation}
Here, we assume that $G_{(\mu A)}=G_{(A\mu )}=0,$ with $A=D+1,...,2D$ and we
identify $x^{A}\rightarrow y^{\mu }$ and $G_{(AB)}\rightarrow G_{(\mu \nu
)}. $ It is not difficult to see that (32) can be rewritten as

\begin{equation}
ds^{2}=G_{(\mu \nu )}^{ab}(x^{\hat{\alpha}})dx_{a}^{\mu }\otimes dx_{b}^{\nu
},  \label{33}
\end{equation}
where $x_{1}^{\mu }\equiv x^{\mu }$ and $x_{2}^{\mu }=y^{\mu }$ and we
assumed that $G_{(\mu \nu )}^{11}=G_{(\mu \nu )}^{22}$ and $G_{(\mu \nu
)}^{12}=G_{(\mu \nu )}^{21}=0.$

On the other hand, if we use the definition

\begin{equation}
z^{\mu }=x^{\mu }+iy^{\mu },  \label{34}
\end{equation}
we find that (32) can be written in the alternative way

\begin{equation}
ds^{2}=G_{(\mu \nu )}(x^{\hat{\alpha}})dz^{\mu }\otimes d\bar{z}^{\nu }.
\label{35}
\end{equation}
In this scenario, since $dx^{\hat{\mu}}\otimes dx^{\hat{\nu}}$ is a
second-rank symmetric tensor the same results follow if we consider the most
general hermitian metric 
\begin{equation}
G_{\hat{\mu}\hat{\nu}}(x^{\hat{\alpha}})=G_{(\hat{\mu}\hat{\nu})}(x^{\hat{%
\alpha}})+iG_{[\hat{\mu}\hat{\nu}]}(x^{\hat{\alpha}}).  \label{36}
\end{equation}
Here, $G_{[\hat{\mu}\hat{\nu}]}$ denotes an antisymmetric tensor metric. Of
course, $G_{\hat{\mu}\hat{\nu}}$ in (36) satisfies the hermitian condition $%
G_{\hat{\mu}\hat{\nu}}=G_{\hat{\mu}\hat{\nu}}^{\dagger }.$

Now, consider the metric $G_{\hat{\mu}\hat{\nu}}(x^{\hat{\alpha}})$ given in
(36), in connection with the exterior product

\begin{equation}
\Omega =\frac{1}{2}G_{\hat{\mu}\hat{\nu}}(x^{\hat{\alpha}})dx^{\hat{\mu}%
}\wedge dx^{\hat{\nu}}.  \label{37}
\end{equation}
Using the exterior product property $dx^{\hat{\mu}}\wedge dx^{\hat{\nu}%
}=-dx^{\hat{\nu}}\wedge dx^{\hat{\mu}},$ we see that (37) leads to

\begin{equation}
\Omega =\frac{i}{2}G_{[\hat{\mu}\hat{\nu}]}(x^{\hat{\alpha}})dx^{\hat{\mu}%
}\wedge dx^{\hat{\nu}}.  \label{38}
\end{equation}
Assuming that $G_{[\mu A]}=G_{[A\mu ]}=0$ we find that (38) becomes 
\begin{equation}
\Omega =\frac{i}{2}G_{[\mu \nu ]}(x^{\hat{\alpha}})dx^{\mu }\wedge dx^{\nu }+%
\frac{i}{2}G_{[AB]}(x^{\hat{\alpha}})dy^{A}\wedge dy^{B}.  \label{39}
\end{equation}
We again make the identification $x^{A}\rightarrow y^{\mu }$ and $%
G_{[AB]}\rightarrow G_{[\mu \nu ]}.$ The formula (39) can be rewritten as

\begin{equation}
\Omega =\frac{i}{2}G_{[\mu \nu ]}(x^{\hat{\alpha}})(dx^{\mu }\wedge dx^{\nu
}+dy^{\mu }\wedge dy^{\nu }).  \label{40}
\end{equation}
Introducing $x_{1}^{\mu }\equiv x^{\mu }$ and $x_{2}^{\mu }=y^{\mu }$ and
assuming that $G_{[\mu \nu ]}^{12}=G_{[\mu \nu ]}^{21}=0$ and $G_{[\mu \nu
]}^{11}=G_{[\mu \nu ]}^{22}\neq 0$ we find that (40) leads to

\begin{equation}
\Omega =\frac{i}{2}G_{[\mu \nu ]}^{ab}(x^{\hat{\alpha}})dx_{a}^{\mu }\wedge
dx_{b}^{\nu }.  \label{41}
\end{equation}
On the other hand, using the definition (34) for $z^{\mu }$ we find that
(40) can also be written as

\begin{equation}
\Omega =\frac{i}{2}G_{[\mu \nu ]}(x^{\hat{\alpha}})dz^{\mu }\wedge d\bar{z}%
^{\nu }.  \label{42}
\end{equation}

Summarizing we have shown that if $G_{(\mu \nu )}^{11}=G_{(\mu \nu
)}^{22}\neq 0$ and $G_{(\mu \nu )}^{12}=G_{(\mu \nu )}^{21}=0,$ and $G_{[\mu
\nu ]}^{11}=G_{[\mu \nu ]}^{22}\neq 0$ and $G_{[\mu \nu ]}^{12}=G_{[\mu \nu
]}^{21}=0$ then

\begin{equation}
ds^{2}=G_{\mu \nu }^{ab}(x_{c}^{\alpha })dx_{a}^{\mu }\otimes dx_{b}^{\nu }
\label{43}
\end{equation}
is equivalent to

\begin{equation}
ds^{2}=G_{\mu \nu }(z^{\alpha },\bar{z}^{\beta })dz^{\mu }\otimes d\bar{z}%
^{\nu },  \label{44}
\end{equation}
and that

\begin{equation}
\Omega =\frac{i}{2}G_{\mu \nu }^{ab}(x_{c}^{\alpha })dx_{a}^{\mu }\wedge
dx_{b}^{\nu }  \label{45}
\end{equation}
is equivalent to

\begin{equation}
\Omega =\frac{i}{2}G_{\mu \nu }(z^{\alpha },\bar{z}^{\beta })dz^{\mu }\wedge
d\bar{z}^{\nu }.  \label{46}
\end{equation}
We recognize in (44) and (46) the formulae used to define the Kahler metric
which in addition satisfies the condition $d\Omega =0$. Therefore we have
shown that under certain anzats the metric $G_{\mu \nu }^{ab}(x_{c}^{\alpha
})$ can be identified with the Kahler metric. This shows that it makes
mathematical sense to consider a metric of the form $G_{\mu \nu
}^{ab}(x_{c}^{\alpha }).$

Our goal is now to use the metric $G_{\mu \nu }^{ab}(x_{c}^{\alpha })$ in
connection with string theory. We find that there are at least two different
ways to achieve this. In fact, in the first case we have the action

\begin{equation}
S_{1}=\frac{1}{2}\int d^{2}\xi \sqrt{-g}g^{ab}(\xi )\partial _{a}x_{c}^{\mu
}\partial _{b}x_{d}^{\nu }G_{\mu \nu }^{cd}(x),  \label{47}
\end{equation}
while in the second case we have$^{42}$

\begin{equation}
S_{2}=\frac{1}{2}\int d^{2}\xi G_{\mu \nu }^{ab}(\xi ,x)\partial _{a}x^{\mu
}\partial _{b}x^{\nu }.  \label{48}
\end{equation}
For our purpose to relate matroid theory with string theory both
possibilities look attractive. The action $S_{1}$ may be useful to
understand T-duality or S-duality because of its property of being symmetric
under the interchange of coordinates $x\leftrightarrow y.$ However, $S_{2}$
is closer to our idea of unified worldsheet-target spacetime metrics when
matter and ghost mix.

In fact, in the particular case 
\begin{equation}
G_{\mu \nu }^{ab}=\sqrt{g}g^{ab}G_{\mu \nu },  \label{49}
\end{equation}
one sees that $S_{2}$ is reduced to the Polyakov action (30). This shows
that ordinary bosonic string theory is contained in a theory associated to
(48). Another particular case for $G_{\mu \nu }^{ab}$ is

\begin{equation}
G_{\mu \nu }^{ab}=\sqrt{g}g^{ab}G_{\mu \nu }+i\varepsilon ^{ab}B_{\mu \nu },
\label{50}
\end{equation}
where $B_{\mu \nu }=-B_{\nu \mu }$ is a two form and $\varepsilon ^{ab}$ is
the completely antisymmetric tensor with $\varepsilon ^{12}=1.$ The choice
(50) leads to a generalized bosonic string theory, the so called nonlinear
sigmal model in two dimensions, in which the string propagates in a
background determined not only by gravity but by the antisymmetric two form
gauge field $B$ with associated field strength $H=dB.$ Finally, the third
example is provided precisely for what we have already discussed when $%
G_{\mu \nu }^{ab}$ is identified with a Kahler metric.

Here, we are not particularly interested in developing the full theory
implied by $S_{2},$ but to point out how $S_{2}$ can be related to matroid
theory via Chern-Simons theory. For this purpose it seems to be convenient
to start by recalling briefly how the Kahler structure is related to
Chern-Simons theory.

There are a number of restrictions which a background field must satisfy in
order to have a consistent string theory. Perhaps one of the most important
is the anomaly cancellation fixed by the constraint$^{4}$

\begin{equation}
dH=trR\wedge R-trF\wedge F,  \label{51}
\end{equation}
where $R$ is the curvature associated to $G_{\mu \nu }$. The formula (51) is
an important restriction for the possible compactifications. One of the most
attractive solutions of (51) is when the ten dimensional space-time vacuum
state is given by T$^{4}\times K$, where T$^{4}$ is a maximally symmetric
four dimensional spacetime and $K$ is a six dimensional Kahler manifold.

Now, it is known that a Kahler metric determines a Kahler manifold, so
Kahler metric is related to string theory through (51). In turn (51)
contains the second Chern class $\int trF\wedge F$ which reduces to the
Chern-Simons form. Therefore, Kahler metric is closely related to
Chern-Simons formalism in string theory. Consequently, the action $S_{2}$
with the choice of $G_{\mu \nu }^{ab}$ as a Kahler metric establishes a
connection between matroid theory and string theory via Chern-Simons
formalism.

\bigskip\ \smallskip\ 

\noindent \textbf{6. ALTERNATIVE CONNECTION BETWEEN MATROID THEORY AND
STRING THEORY}

\smallskip\ 

The observations in the previous section may motivate to look for a
Chern-Simons formulation for strings. An attempt in this direction has been
proposed by Zaikov.$^{18}$ Dolan and Tchrakian$^{43}$ have proposed a
similar structure which has been extensively studied by Castro.$^{44}$ One
of the roots of these developments is the Schild type construction for
strings. In this section we shall show that this kind of construction offers
a more direct connection between matroid theory and string theory.

Let us first recall the Schild type construction for strings. It is well
known that from the action (30) one can derive the Naumbu-Goto action

\begin{equation}
S=T\int d^{2}\xi \sqrt{-h},  \label{52}
\end{equation}
where $h$ is the determinant of

\begin{equation}
h_{ab}=\partial _{a}x^{\mu }\partial _{b}x^{\nu }G_{\mu \nu }(x).  \label{53}
\end{equation}
Here, we have restored the tension $T$ of the string.

It is not difficult to see that $h$ can be written as

\begin{equation}
h=\sigma ^{\mu \nu }\sigma _{\mu \nu },  \label{54}
\end{equation}
where

\begin{equation}
\sigma ^{\mu \nu }=\frac{1}{[2!]^{\frac{1}{2}}}\varepsilon ^{ab}v_{a}^{\mu
}(\xi )v_{b}^{\nu }(\xi ).  \label{55}
\end{equation}
Here, $v_{a}^{\mu }(\xi )$ is defined by

\begin{equation}
v_{a}^{\mu }(\xi )=\partial _{b}x^{\mu }(\xi ).  \label{56}
\end{equation}

One can show that the action (52) is equivalent to

\begin{equation}
S_{p}^{(1)}=\int d^{2}\xi (\sigma ^{\mu \nu }p_{\mu \nu }-\frac{\gamma }{2}%
(p^{\mu \nu }p_{\mu \nu }+T^{2})),  \label{57}
\end{equation}
where $\gamma $ is a lagrange multiplier and $p_{\mu \nu }$ is a linear
momentum associated to $\sigma ^{\mu \nu }$. If we eliminate $p_{\mu \nu }$
from this action we get

\begin{equation}
S_{p}^{(1)}=\frac{1}{2}\int d^{2}\xi (\gamma ^{-1}\sigma ^{\mu \nu }\sigma
_{\mu \nu }-\gamma T).  \label{58}
\end{equation}
By eliminating $\gamma $ from (58) we recover action (52). The importance of
(57) or (58) is that it now makes sense to set $T=0.$ In this case (58) is
reduced to the Schild type null 1-brane action$^{13}$. Here, we are
interested in connecting (58) with matroid theory.

Observe that we can relate (58) with matroid theory if we connect $\sigma
^{\mu \nu }$ with a matroid structure. Therefore, the question is how
expression (55) is related to matroids.

Consider the matrix

\begin{equation}
D=\left( 
\begin{array}{cccc}
1 & 0 & 1 & 1 \\ 
0 & 1 & 1 & -1
\end{array}
\right) .  \label{59}
\end{equation}
The matrix $D$ is a realization of the matroid $E=\{1,2,3,4\}$ and

\begin{equation}
\mathcal{B}=\{\{1,2\},\{1,3\},\{1,4\},\{2,3\},\{2,4\},\{3,4\}\}.  \label{60}
\end{equation}
This is the uniform matroid $U_{2,4}.$ The elements of $E$ are identified
with the columns of $D$ in the form $1\rightarrow b_{a}^{1},$ $2\rightarrow
b_{a}^{2}$, $3\rightarrow b_{a}^{3}$ and $4\rightarrow b_{a}^{4}$ where $%
b_{a}^{\mu }$ are the four columns of $D$ and the index $a$ runs from $1$ to 
$2$. In this case the chirotope formula (9) reads as

\begin{equation}
\chi (\mu ,\nu )\equiv sign\det (\mathbf{b}^{\mu },\mathbf{b}^{\nu })\in
\{-1,0,1\}.  \label{61}
\end{equation}
In tensor notation we can write (61) in the form

\begin{equation}
\chi (\mu ,\nu )\equiv sign(\varepsilon ^{ab}b_{a}^{\mu }b_{a}^{\nu })\in
\{-1,0,1\}  \label{62}
\end{equation}
Let us define

\begin{equation}
\Sigma ^{\mu \nu }\equiv \varepsilon ^{ab}b_{a}^{\mu }b_{a}^{\nu }
\label{63}
\end{equation}
so that $\chi (\mu ,\nu )\equiv sign\Sigma ^{\mu \nu }$. Comparing (55) and
(63) we observe the great similarity between the two formulae. The main
difference is the local property of (55). This is in a certain sense similar
to the relation between the flat Minkowski metric $\eta _{\mu \nu
}=diag(1,...,1)$ and a curved metric $g_{\mu \nu }=g_{\mu \nu }(x).$ Let us
discuss this analogy in more detail.

Consider a $n-$dimensional manifold $M.$ A tangent bundle associated to $M$
can be defined as

\begin{equation}
TM=\bigcup_{x\in M}T_{x}(M),  \label{64}
\end{equation}
where $T_{x}(M)$ is the tangent space attached at each point $x$ of $M$. The
cotangent bundle

\begin{equation}
T^{\ast }M=\bigcup_{x\in M}T_{x}^{\ast }(M)  \label{65}
\end{equation}
is defined through the isomorphism $w:T_{x}(M)\rightarrow R$, where $w\in
T_{x}^{\ast }(M)$. The curved metric $g_{\mu \nu }(x)$ can be understood as
the symmetric positive definite map

\begin{equation}
g:T_{x}(M)\rightarrow T_{x}^{\ast }(M).  \label{66}
\end{equation}
In tensor notation (45) becomes

\begin{equation}
a_{\mu }=g_{\mu \nu }a^{\nu },  \label{67}
\end{equation}
where $a^{\nu }\in T_{x}(M)$ and $a_{\mu }\in T_{x}^{\ast }(M).$ A tangent
bundle is an example of fiber bundles. In this case the fiber is the tangent
space $T_{x}(M)$.

Therefore, the transition from $\eta _{\mu \nu }$ to $g_{\mu \nu }(x)$ is
determined by the transition from flat vector space to a fiber bundle.
Similarly, we may say that the transition from $\Sigma ^{\mu \nu }$ to $%
\sigma ^{\mu \nu }(\xi )$ is determined by the transition form matroid
structure to matroid bundle structure. But the question is now to understand
what we mean by matroid bundle. Fortunately the mathematicians have already
developed such a concept.$^{14-17}$ The matroid bundle concept can be
understood as the combinatorial analogue of a fiber bundle. In a matroid
bundle the fiber of a bundle is replaced by a matroid. It is worth
mentioning that recently, matroid bundle structure was used to combine the
concepts of matroid and gravity in a proposed theory called gravitoid.$^{29}$

Summarizing, we may start with a rank-two realizable matroid $M$ such as the
example in (59). We construct the chirotope $\chi (\mu ,\nu )$ by means of
the tensor $\Sigma ^{\mu \nu }$ as in the formula (63). We then make the
transition from a matroid structure to matroid bundle structure in such a
way that $\Sigma ^{\mu \nu }\rightarrow \sigma ^{\mu \nu }(\xi )$. If we can
go from (55) to (56) our task is finished.

Consider the object

\begin{equation}
F_{ab}^{\mu }=\partial _{a}v_{b}^{\mu }(\xi )-\partial _{b}v_{a}^{\nu }(\xi
).  \label{68}
\end{equation}
If $F_{ab}^{\mu }$ vanishes then a solution of (68) is $v_{a}^{\mu }(\xi )=%
\frac{\partial x^{\mu }}{\partial \xi ^{a}}$ where $x^{\mu }$ is in this
context a gauge function. In this case, one says that $v_{a}^{\mu }(\xi )$
is a pure gauge. This kind of scenario can be derived from an abelian
Chern-Simons action for $F_{ab}^{\mu }$ in the form

\begin{equation}
S_{CS}=\frac{k}{2\pi }\int d^{3}\xi \varepsilon ^{ijk}v_{i}^{\mu }F_{jk\mu }.
\label{69}
\end{equation}
This shows once again the great importance of Chern-Simons formulation for a
matroid/string connection. In fact, if we substitute expression (56) into
(69) then the Zaikov%
\'{}%
s Chern-Simons type action for strings is obtained.

It is worth mentioning that in general a p-form $F$ that can be written as $%
F=v_{1}\wedge ...\wedge v_{p}$ for some $v_{1},...,v_{p}\in R^{n}$ is called
decomposable. This means that the two form $\sigma ^{\mu \nu }(\xi )$ given
in (55) is decomposable. It turns out that decomposable forms may be
considered as the starting point to construct the realization space of an
oriented matroid (see Ref. 26 chapter 8) These observations provide an
additional evidence for the close relation between $\sigma ^{\mu \nu }(\xi )$
and oriented matroids.

\bigskip\ \smallskip\ 

\noindent \textbf{7. COMMENTS}

\smallskip\ 

In the present work we have shown that the Kahler metric may provide an
important bridge to connect matroid theory and string theory. Specifically
using a generalized string theory we established the following
identifications: $string-theory\leftrightarrow Kahler-structure$ ; $%
Chern-Simons-theory\leftrightarrow Kahler-structure$ and $%
matroid-theory\leftrightarrow Chern-Simons-theory$. Moreover, it is natural
to expect that this kind of relations suggest a direct link between the
generalized string theory described by (48), which by convenience we shall
call Kahler-string action, and the pure Chern-Simons theory. But at first
sight it is improbable that this more direct relation exists. The reason is
that pure Chern-Simons theory does not provide us enough degrees of freedom
to describe the dynamics even for the heterotic string theory. Therefore,
these observations suggest that there must exist a generalized Chern-Simons
action which is reduced to the Kahler-string action (48). Of course, the
Kahler structure must play an important role in this generalized
Chern-Simons theory. Happily, such a theory along this idea has already been
proposed. In fact, some years ago Nair and Schiff$^{45}$ proposed what they
called Kahler-Chern-Simons theory. The action proposed by Nair and Schiff
has the form

\begin{equation}
S_{KCS}=\frac{k}{2\pi }\int_{M^{3}\times R}Tr(A\wedge dA+\frac{2}{3}A\wedge
A\wedge A)\Omega ,+Tr(F\Phi +\Phi F),  \label{70}
\end{equation}
where $F$ is the field strength in $B^{4}\times R$, $\Phi $ is a Lie algebra
value $(2,0)$ form and $M^{4}$is a Kahler manifold of real dimension four.
It turns out that $S_{KCS}$ provides an action description of antiself-dual
gauge fields (instantons) on a four dimensional Kahler manifolds. Here, we
are interested in seeing if under quantization $S_{KCS}$ is reduced to the
Kahler-string action (48). It can be shown, however, that up to WSW term, $%
S_{KCS}$ leads to the action

\begin{equation}
S_{3}=\frac{1}{2}\int d^{2+2}\xi \sqrt{-g}g_{ij}^{ab}(\xi )\frac{\partial
x^{\mu }}{\partial \xi _{i}^{a}}\frac{\partial x^{\nu }}{\partial \xi
_{i}^{a}}G_{\mu \nu }(x),  \label{71}
\end{equation}
rather than (48). Here, $g_{ij}^{ab}$ can be identified with a Kahler metric
on $B^{4},$ while $G_{\mu \nu }(x)$ is given by $G_{\mu \nu }(x)=\partial
_{\mu }U\partial _{\nu }U^{-1},$ where $U$ is a locally defined $G$-valued
function related to the gauge field $A$ by $A=U^{-1}dU.$ Therefore, we
conjecture that there must be a slightly different action from (70), with
the property of reducing to (48).

Nevertheless, the action $S_{KCS}$ may be of special interest to relate
matroid theory not only to string theory but to M-theory itself. In fact, it
is known that $S_{KCS}$ leads to a theory in terms of fields in a target
space of $N=2$ strings.$^{46}$ In turn, $N=2$ strings is one of the main
proposals of M-theory.$^{47}$ Moreover, it has been pointed out in Ref. 48
that a number of similarities exist between the other main proposal of
M-theory, namely Matrix theory.$^{49}$ In principle, for $S_{KCS}$ one may
repeat the formalism of section 4. In fact, consider the partition function

\begin{equation}
Z_{KCS}(L,k)=\int DA\exp (S_{KCS})\prod\limits_{r=1}^{n}W(L_{r},\rho _{r}),
\label{72}
\end{equation}
where $W(L_{i},\rho _{i})$ are the Wilson lines defined in (14). It is
tempting to speculate that $Z_{KCS}(L,k)$ must be related to some knot
invariant in a similar way that $Z_{CS}(L,k)$ is related to the Jones
polynomials. From this idea, since knots are in one to one correspondence
with signed graphs one should expect to find the desired relation between
matroid theory and $Z_{KCS}(L,k)$, which may lead eventually to a matroid
theory and M-theory connection.

In section 6, we discuss an alternative possibility to connect matroids and
strings. The idea is to consider the Schild type action for strings. (This
action is equivalent to the Nambu-Goto action and may be considered as the
starting point for the Zaikov and Dolan-Tchrakian constructions for
p-branes.) We identify the $\sigma ^{\mu \nu }$ factor in (55) with the
tensor $\Sigma ^{\mu \nu }$ associated to the chirotope $\chi (\mu ,\nu )$
via the matroid bundle notion. Such identification may lead to the
matroid/string connection if one consider the Chern-Simons structure in
(69), which becomes a Zaikov type action after using (56). It is important
to mention that the discussion of section 6 can be generalized in
straightforward way to any p-brane$^{50}.$

Another possible route to connect matroid theory with M-theory comes from
the work of Gopakumar and Vafa$^{51-52}$ who have proved that topological
strings are closely related to M- theory. Since Chern-Simons formalism is
linked to topological strings,$^{35}$ it seems that we are closer to make
the matroid theory the underlying mathematical structure of M-theory (see
Ref. 53 for interesting observations about M- theory).

Finally, besides the possible connection between matroid theory and string
theory the present formalism may be of special interest for quantum gravity
based on the Ashtekar formalism$^{54}$ (see also Ref. 55 and references
therein). The most interesting solutions of the Ashtekar constraints
correspond to the Witten's partition function. Consequently, the duality
symmetries (24) may also play an important role in such solutions. It is
known that the Vasiliev invariants become an important tool in the loop
solutions of quantum canonical gravity in the Ashtekar formalism. Since the
Vasiliev invariants can be understood as a generalization of the Jones
polynomials, it may be interesting for further research to investigate
whether matroid theory can be connected to such invariants.

\bigskip

\noindent \textbf{Acknowledgment:} I would like to thank M. C. Mar\'{i}n, E.
Fr\'{i}as, E. Ley-Koo and G. Campoy for helpful comments.

\end{document}